\documentclass[12pt]{article}

\def\hybrid{\topmargin 0pt      \oddsidemargin 0pt
        \headheight 0pt \headsep 0pt
        \voffset=-0.5cm
        \textwidth 6.5in        
        \textheight 9in         
        \marginparwidth 0.0in
        \parskip 5pt plus 1pt   \jot = 1.5ex}
\catcode`\@=11
\def\marginnote#1{}

\newcount\hour
\newcount\minute
\newtoks\amorpm
\hour=\time\divide\hour by60
\minute=\time{\multiply\hour by60 \global\advance\minute by-\hour}
\edef\standardtime{{\ifnum\hour<12 \global\amorpm={am}%
        \else\global\amorpm={pm}\advance\hour by-12 \fi
        \ifnum\hour=0 \hour=12 \fi
        \number\hour:\ifnum\minute<10 0\fi\number\minute\the\amorpm}}
\edef\militarytime{\number\hour:\ifnum\minute<10 0\fi\number\minute}

\def\draftlabel#1{{\@bsphack\if@filesw {\let\thepage\relax
   \xdef\@gtempa{\write\@auxout{\string
      \newlabel{#1}{{\@currentlabel}{\thepage}}}}}\@gtempa
   \if@nobreak \ifvmode\nobreak\fi\fi\fi\@esphack}
        \gdef\@eqnlabel{#1}}
\def\@eqnlabel{}
\def\@vacuum{}
\def\draftmarginnote#1{\marginpar{\raggedright\scriptsize\tt#1}}

\def\draftlabel#1{{\@bsphack\if@filesw {\let\thepage\relax
   \xdef\@gtempa{\write\@auxout{\string
      \newlabel{#1}{{\@currentlabel}{\thepage}}}}}\@gtempa
   \if@nobreak \ifvmode\nobreak\fi\fi\fi\@esphack}
        \gdef\@eqnlabel{#1}}
\def\@eqnlabel{}
\def\@vacuum{}
\def\draftmarginnote#1{\marginpar{\raggedright\scriptsize\tt#1}}

\def\draft{\oddsidemargin -.5truein
        \def\@oddfoot{\sl preliminary draft \hfil
        \rm\thepage\hfil\sl\today\quad\militarytime}
        \let\@evenfoot\@oddfoot \overfullrule 3pt
        \let\label=\draftlabel
        \let\marginnote=\draftmarginnote
   \def\@eqnnum{(\theequation)\rlap{\kern\marginparsep\tt\@eqnlabel}%
\global\let\@eqnlabel\@vacuum}  }

\def\numberbysection{\@addtoreset{equation}{section}
        \def\theequation{\thesection.\arabic{equation}}}

\def\underline#1{\relax\ifmmode\@@underline#1\else
        $\@@underline{\hbox{#1}}$\relax\fi}

\def\titlepage{\@restonecolfalse\if@twocolumn\@restonecoltrue\onecolumn
     \else \newpage \fi \thispagestyle{empty}\c@page\z@
        \def\thefootnote{\fnsymbol{footnote}} }

\def\endtitlepage{\if@restonecol\twocolumn \else  \fi
        \def\thefootnote{\arabic{footnote}}
        \setcounter{footnote}{0}}  
\relax

\numberbysection
\hybrid

\def\beq{\begin{equation}}
\def\eeq{\end{equation}}
\def\p{\partial}
\def\G{\Gamma}
\def\g{\gamma}
\def\s{\sigma}

\def\L{{\cal L}}

\def\a{\alpha}
\def\b{\beta}

\def\l{\lambda}

\def\A{{\cal A}}

\def\V{{\cal V}}
\def\D{{\cal D}}
\def\F{{\cal F}}

\def\L{{\cal L}}
\def\K{{\cal K}}

\def\LD{\L^{D_+,\,D_-}_{\g(n),\,\a(n)}}
\def\LL{\L^{D_+,\,D_-}}
\def\LLL{\L^{D_+,\,D_-}_{N,\, C,\,  \Delta}}
\def\LO{\L^{D_+,\,D_-}_{N,\, 0,\,  \Delta}}
\def\LT{\hat \L_{N,\, \Delta}}
\def\M{{\cal M}}

\def\P{{\cal P}}
\def\X{{\cal X}}
\def\SP{{\cal S}}

\def\dim{{\rm dim}}
\def\res{{\rm res}}
\def\F{{\cal F}}

\def\wh{\widehat}

\def \matrix #1 {\left(\begin{array}{cc} #1 \end{array}\right)}
\newtheorem{theo}{Theorem}[section]

\newtheorem{cor}{Corollary}[section]
\newtheorem{lem}{Lemma}[section]

\begin{document}

\input epsf

\begin{titlepage}
\title{Integrable chains on algebraic curves}

\author{I.Krichever \thanks{Columbia University, New York, USA and
Landau Institute for Theoretical Physics and ITEP, Moscow, Russia; e-mail:
krichev@math.columbia.edu. Research is supported in part by National Science
Foundation under the grant DMS-01-04621.}}

\date{August, 2003}

\maketitle

\begin{abstract}
The discrete Lax operators with the spectral parameter on an algebraic
curve are defined. A hierarchy of commuting flows on the space
of such operators is constructed. It is shown that these flows are
linearized by the spectral transform and can be explicitly solved in
terms of the theta-functions of the spectral curves.
The Hamiltonian theory of the corresponding systems is analyzed.
The new type of completely integrable Hamiltonian systems associated with
the space of rank $r=2$ discrete Lax operators on a {\it variable} base curve
is found.
\end{abstract}

\end{titlepage}

\section{Introduction.}
The main goal of this work is to construct discrete analogs of the
the zero-curvature equations on an algebraic curve introduced in \cite{kn1}, and
identified later in \cite{kr-hit} with infinite-dimensional
field analogs of the Hitchin system \cite{hit}. The starting point
was an attempt to find the general setting for the difference-differential
equations introduced in the recent work \cite{kn2}
by S.Novikov and the author. These equations were used for the construction of
high rank solutions of the $2D$ Toda lattice equations simultaneously with
the construction of commuting high rank difference operators.

Almost all (1+1)-soliton equations admit  zero-curvature
representation (\cite{zakh})
\beq\label{curv0}
\p_t L-\p_x M+[L,M]=0,
\eeq
where $L(x,t,z)$ and $M(x,t,z)$ are {\it rational} matrix functions of
the {\it spectral} parameter $z$. The discrete analog of (\ref{curv0}) is
the equation
\beq\label{dcurv}
\p_t L_n=M_{n+1}L_n-L_nM_n,
\eeq
where, as before, $L_n=L_n(t,z)$ and $M_n=M_n(t,z)$ are rational functions
of the spectral parameter. In both the cases poles of $L$ and $M$ are fixed.
The singular parts of $L$ and $M$ at the poles are dynamical variables. Their number
equals the number of equations equivalent to (\ref{curv0}) or (\ref{dcurv}),
respectively.

The Riemann-Roch theorem implies that naive direct generalization of
equations (\ref{curv0}, \ref{dcurv}) for matrix functions, which are meromorphic
on an algebraic curve $\G$ of genus $g>0$, leads to an over-determined system of
equations (see details in \cite{kr-hit}). In \cite{kn1} it was found
for the continuous case, that if the matrix functions $L$ and $M$ have
moving poles with special dependence on $x$ and $t$ besides fixed poles,
then equation (\ref{curv0}) is a well-defined system on the space of singular parts
of $L$ and $M$ at fixed poles. In \cite{kr-hit} it was shown that a part of these
equations can be used to express $M$ in terms of $L$. After that the zero-curvature
equations can be seen as a hierarchy of commuting flows on the space of admissible
matrix-valued meromorphic functions $L(x,z),\ z\in \G $.
The admissible meromorphic matrix-functions on a smooth genus $g$ algebraic curve
were identified with $x$-connections on $x$-parametric families $\V(x)$
of stable rank $r$ and degree $rg$ holomorphic vector bundles on $\G$.
In the stationary case, the factor-space $\L^{\K}/SL_r$
of $x$-independent connections $L(z)$ with the divisor of poles equivalent to
the canonical divisor $\K$ is isomorphic to the phase space of the Hitchin
system. The latter is the cotangent space $T^*(\M)$ of the moduli space $\M$
of stable rank $r$ and degree $rg$ holomorphic vector-bundles on $\G$.

A discrete analog of $x$-parametric family of vector bundles is a sequence
of vector bundles $\V_n\in \M$. The discrete analog of a meromorhic $x$-connection
with the pole divisor $D_+$ is a chain $L_n(z)$ of meromorphic homomorphisms
$L_n\in H^{0}({\rm Hom}(\V_{n+1}, \V_{n}(D_+))).$
It is assumed that $L_n$ is almost everywhere invertible and the inverse
homomorphism has a fixed divisor of poles $D_-$, i.e.
$L_n^{-1}\in H^{0}({\rm Hom}(\V_{n}, \V_{n+1}(D_-)))$.
In the next section we show the algebraic integrability
of the space $\LL_N$ of periodic chains, considered modulo gauge
transformations $L_n'=g_{n+1}L_n g_n, \ g_n\in GL_r$. Namely, we show that an open
set of the factor-space $\LL_N/GL_r^N$ is isomorphic to an open set of
Jacobian bundle over the space $\SP^{D_+,\, D_-}$ of the spectral curves $\hat \G$.
The spectral curves are defined by the characteristic equation for
the monodromy operator $T=L_N\,L_{N-1}\cdots L_1$, which
represents them as an $r$-sheet cover of the base curve $\pi:\hat \G\longmapsto \G$.
The spectral transform identifies an open set of the {\it restricted} chains,
corresponding to the sequences of bundles $\V_n$ with fixed determinant,
with an open set of the bundle over $\SP^{D_+,\, D_-}$
with fibers $J_C(\hat \G)$, which are preimages in $J(\hat\G)$ of some
point of $J(\G)$.

In Section 3 a hierarchy of gauge invariant commuting Lax equations
(\ref{dcurv}) is constructed. For periodic chains these equations are linearized
by the spectral transform and can be explicitly solved in terms of
the Riemann theta-function of the spectral curve.

One of the most general approaches to the Hamiltonian theory of the soliton
equations, admitting the zero-curvature representation with the spectral parameter on
the rational or an elliptic curve, is based on the classical $r$-matrix description
of the Poisson structure of the corresponding phase space. Essentially there are two
types of the Poisson structures reflecting Lie algebraic or Lie group nature of the
auxiliary linear problem. The first type usually is referred  to as
linear brackets and the second one as quadratic or Sklyanin brackets
(see the book \cite{ft}, survey \cite{rs} and references therein).

The alternative approach to the Hamiltonian theory of the soliton
equations was developed in \cite{kp1,kp2}. It is based on the existence of some
universal two-form defined on a space of meromorphic matrix-functions. A direct and
simple corollary of the definition of this form is that its contraction by the
vector-field, defined by zero-curvature equation, is an {\it exact} one-form.
Therefore, whenever this form is non-degenerate the corresponding system is
the Hamiltonian system. In Section 4 it shown that the group version of
our approach is applicable to the case of the Lax chains on algebraic curves
of arbitrary genus (see \cite{kp3} for the genus zero case). It is shown that
for the cases of the rational and ellitic
curves the corresponding form $\omega$ restricted to the symplectic leaves
$\P_*\subset \LL_N/GL_r^N$ is induced by the Sklyanin symplectic form on the
space of the monodromy matrices. For $g>1$ the form
$\omega$ is degenerate on $\P_*$.
That does not allow us to treat Lax equations (\ref{dcurv}) for $g>1$
within the framework of conventional Hamiltonian theory.
At the same time the corresponding system has new very unusual features.
The space $\P_*$ is equipped by $g$-parametric {\it family}
of two-forms $\omega_{dz}$, parameterized by the holomorphic differentials
$dz$ on $\G$. For all of them the fibers
$J_C(\hat \G)$ of the spectral bundle are maximum isotropic subspaces.
For each of the flows defined by the Lax equations on $\P_*$, the contraction
$i_{\p_t}\omega_{dz}$ is an exact form $ \delta H^{dz}$.
Although each of the forms $\omega_{dz}$ is degenerate on $\P_*$, their
family is {\it non-degenerate}.

In a certain sense the state-of-art
described above is dual to that in the theory of bi-Hamiltonian systems.
Usually the bi-Hamiltonian structure is defined on the Poisson manifold
equipped by a family of compatible brackets. The vector-fields
that are Hamiltonian with respect to one bracket are Hamiltonian
with respect to the other ones, but correspond to different Hamiltonians.
The drastic difference between the bi-Hamiltonian systems and the systems
$\P_*$ of restricted chains is in the nature of the symplectic leaves.
For the bi-Hamiltonian systems usually they are {\it globally} defined as levels
of single-valued action-type variables.
For $\P_*$ the form $\omega_{dz}$ becomes
non-degenerate on levels of {\it multi-valued} angle-type functions.

It is worth to understand if there exists the general Hamiltonian-type setting,
in which these characteristic features of $\P_*$ for $g>1$ provide the basis for
something that might be the notion of {\it super-integrable} systems.
It is also possible, that there is no need for the new setting.
The results of the last section provide some evidence that the Lax chains
on the fixed base curve $\G$ might be "extended" to the
conventional completely integrable Hamiltonian system.
Namely, we show that for the rank $r=2$ the space of the periodic Lax chains with
{\it variable base curve $\G$} is the Poisson
manifold with leaves $\hat \P_{\Delta}$, corresponding to the chains
(modulo gauge equivalence) with fixed determinant $\Delta$ of the monodromy matrix
$T$, and with the fixed regular eigenvalue $w$ of $T$ at the punctures
$P_k^{\pm}$. The universal form defines the structure of a
completely integrable system on $\hat \P_{\Delta}$. The Hamiltonians of the
Lax equations on
$\hat \P_{\Delta}$ are in involution. They are given by the formula
$$ H_f=\sum_{q=\hat \P_k^{\pm}} \res_q \, (f\ln w) d\ln \Delta,$$
where $f$ is a meromorphic function on $\G$ with poles at the punctures
$P_k^{\pm}$. The common level of all the integrals $H_f$ is identified with
the Prim variety of the corresponding spectral curve.

\section{Algebraic integrability of chains}

Let $\G$ be a smooth genus $g$ algebraic curve. According to \cite{tyur},
a generic stable, rank $r$ and degree $rg$ holomorphic vector bundle
$\V$ on $\G$ is parameterized by a set of $rg$ distinct points $\g_s$ on $\G$,
and a set of $r$-dimensional vectors $\a_s=(\a_s^i),$ considered modulo scalar
factor $\a_s\to \lambda_s \a_s$ and
a common gauge transformation $\a_s^i\to g_j^i \a_s^j,\ \ g\in GL_r$, i.e. by a
point of the factor-space
$$\M_0=S^{rg} \left(\G\times CP^{r-1}\right)/GL_r$$
In \cite{kn1,kr2} the data $(\g,\a)=(\g_s,\a_s), \ s=1,\ldots,rg, \ i=1,\ldots,r,$
were called the Tyurin parameters.

Let $D_{\pm}$ be two effective divisors on $\G$ of the same degree $\D$.
Throughout the paper, if it is not stated otherwise, it is assumed
that all the points of the
divisors $D_{\pm}=\sum_{k=1}^{\D} P_k^{\pm}$ have multiplicity $1$,
$P_k^{\pm}\neq P_m^{\pm},
k\neq m$.
For any sequence of the Tyurin parameters
$(\g(n),\a(n))$ we introduce the space $\L^{D_+,\,D_-}_{\g(n),\,\a(n)}$
of meromorphic matrix functions $L_n(q), q\in \G,$ such that:

$1^0.$ $L_n$ is holomorphic except at the points $\g_s$, and at the  points $P_i^+$
of $D_+$, where it has at most simple poles;

$2^0$. the singular coefficient $L_{s}(n)$ of the Laurent expansion
of $L_n$ at $\g_s$
\beq\label{Ls}
L_n(z)={L_{s}(n)\over z-z_s}+O(1),\ \ z_s=z(\g_s),
\eeq
is a rank 1  matrix of the form
\beq \label{Ls0}
L_{s}(n)=\b_s(n)\a_s^T(n)\ \longleftrightarrow L_{s}^{ij}(n)=\b_s^i(n)\a_s^j(n),
\eeq
where $\b_s(n)$ is a vector, and
$z$ is a local coordinate in the neighborhood of $\g_s$;

$3^0$. the vector $\a_s^T(n+1)$ is a left null-vector of the evaluation of $L_n$
at $\g_s(n+1)$, i.e.
\beq\label{Ls1}
\a_s(n+1)L_n(\g_s(n+1))=0;
\eeq

$4^0$. the determinant of $L_n(q)$ has simple poles at the points $P_k^+, \g_s(n)$,
and simple zeros at the points $P_k^-,\g_s(n+1)$.

The last condition implies the following constraint for the equivalence classes
of the divisors
\beq\label{class}
[D_+]-[D_-]=\sum_s \left[\g_s(n+1)-\g_s(n)\right]\in J(\G),
\eeq
where $J(\G)$ is the Jacobian of $\G$.
If $2N>g(r+1)$, then the Riemann-Roch theorem and
simple counting of the number of the constraints (\ref{Ls})-(\ref{Ls1})
imply that the functional dimension of $\LD$ (its dimension as the space of
functions of the discrete variable $n$) equals
$2\D(r-1)-gr^2+g+r^2$.

The geometric interpretation of $\LD$ is as follows.
In the neighborhood of $\g_s$ the space of local sections
of the vector bundle $V_{\g,\a}$, corresponding
to $(\g,\a)$, is the space $\F_s$ of meromorphic functions
having a simple pole at $\g_s$ of the form
\beq\label{F}
f(z)={\l_s\a_s^T\over z-z(\g_s)}+O(1),\ \ \l_s\in C\, .
\eeq
Therefore, if $\V_n$ is a sequence of the vector bundles on $\G$, corresponding
to the sequence of the Tyurin parameters $(\g(n),\a(n))$, then $L_n$ can be
seen as a homomorphism of the vector bundle $\V_{n+1}n$ to the vector bundle
$\V_{n}(D_+)$, obtained from $\V_{n}$ with the
help of simple Hecke modification at the punctures $P_k^+$, i.e.
\beq\label{maps}
L_n\in {\rm Hom} (\V_{n+1},\V_{n}(D_+)).
\eeq
These homomorphisms are invertible almost everywhere. The inverse matrix-functions
are the homomorphisms of the vector bundles
\beq\label{maps-}
L_n^{-1}\in {\rm Hom} (\V_{n},\V_{n+1}(D_-)).
\eeq
The total space $\LL$ of the chains corresponding to all the sequences of the
Tyurin parameters is a bundle over the space of sequences of holomorphic vector bundles
\beq\label{tot}
\LL\longmapsto \prod_n \M_0\equiv \{\g(n),\a(n)\}\, .
\eeq
The fibers of this bundle are just the spaces $\LD$.
Our next goal is to show algebraic integrability of the total space $\LL_N$ of
the {\it $N$-periodic} chains,
$L_n=L_{n+N}$. Equation (\ref{class}) implies that the periodicity of chains
requires the following constraint on the equivalence classes of the divisors
$D_{\pm}$:
\beq\label{class1}
N\left([D_+]-[D_-]\right)=0\in J(\G),
\eeq
which will be always assumed below.
The dimension of $\LL_N$ equals
\beq\label{dimL}
\dim \ \LL_N =2N\D(r-1)+Nr^2+g.
\eeq
The last term in the sum corresponds to the dimension of the equivalence class
of the set $\g(0)$, which defines the equivalence
classes of all the divisors $\g(n)$ with the help of (\ref{class}).

Let $L_n\in \LL_N$ be a periodic chain. Then the Floque-Bloch solutions of the
equation
\beq\label{laxL}
\psi_{n+1}=L_n\psi_n
\eeq
are solutions that are eigenfunctions for the monodromy operator
\beq\label{mon}
T_n\psi_n=\psi_{n+N}=w\psi_n,\ \
T_n=L_{n+N-1}\cdots L_{n+1}L_n.
\eeq
The monodromy matrix $T_n(q)$ belongs to the space of the Lax matrices introduced
in \cite{kr-hit}, $T_n\in \L^{ND_+}, \ T_n^{-1}\in\L^{ND_-}$.
The Floque-Bloch solutions are parameterized by the points $Q=(w,q), \ q\in \G,$
of the so-called spectral curve $\hat \G$ defined by the
characteristic equation
\beq\label{char}
R(w,q)=\det\left(w\cdot 1-T_{n_0}(q)\right)=w^r+\sum_{i=0}^{r-1}r_i(q)w^i=0.
\eeq
From (\ref{Ls})-(\ref{Ls1}) it follows that the monodromy matrix $T_{n_0}$
has poles at the punctures $P_k^{+}$ and the points $\g_s(n_0)$.
Therefore, {\it a priopi} the coefficients $r_i(q)$ of (\ref{char})
might have poles only at the same set of points.
The characteristic equation is $n_0$-independent, because
$T_{n_0+1}=L_{n_0+N}T_{n_0}L_{n_0}^{-1}=L_{n_0}T_{n_0}L_{n_0}^{-1}$.
Hence, $r_i(q)$ are meromorphic functions on $\G$ with the poles only at
the punctures $P_k^+$. Equation (\ref{char}) defines an affine part of the
spectral curve. Let us consider its compactification over the punctures $P_k^+$.
By definition $L_n$ and its determinant have simple poles at $P_k^+$. Hence, its
residue at this point has rank 1, i.e. the Laurent expansion of $L_n$ at $P_k^+$
has the form
\beq\label{t1}
L_n={h_k(n)p_k^T(n)\over z-z(P_k^+)}+O(1),
\eeq
where $z$ is a local coordinate in the neighborhood of $P_k^+$, and
$h_k(n),p_k(n)$ are $r$-dimensional vectors. These vectors are defined up
to the gauge transformation
\beq\label{gg}
h_k(n)\longmapsto c_{k,n}q_k(n),\ \
h_k(n)\longmapsto c_{k,n}^{-1}q_k(n),
\eeq
where $c_{k,n}$ is a constant.
The leading term of the Laurent expansion of the monodromy matrix equals
\beq\label{t2}
T_{n_0}=
\prod_{n=n_0}^{n_0+N-1}\left(p_k^T(n+1)h_k(n)\right)
{h_k(n_0+N-1)p_k^T(n_0)\over (z-z(P_k^+))^N}+O\left(\left(z-z(P_k^+)\right)^{-N+1}
\right).
\eeq
Equation (\ref{t2}) implies that in the neighborhood of $P_k^+$
one of the roots of the characteristic equation
has the form
\beq\label{w}
w=(z-z(P_k^+))^{-N}\left(c_k^++O(z-z(P^+_k))\right),\ \
c_k^+=\prod_{n=0}^{N-1}\left(p_k^T(n+1)h_k(n)\right).
\eeq
The corresponding compactification point of $\hat \G$ is smooth, and
will be denoted by $\hat P_k^+$. The determinant of $T_{n_0}$ has the pole of order
$N$ at $P_k^+$.
Therefore, in the general position all the other branches of $w(z)$ are regular
at $P_k^+$. The coefficients $r_i(z)$ are the elementary symmetric polynomials of
the branches of $w(z)$. Hence, all of them have poles at $P_k^+$ of order $N$.
Note, that the coefficient $r_0(z)=\det T_{n_0}$ has zero of order $N$ at $P_k^-$.

The same arguments applied to $L_n^{-1}$ show that
over the puncture $P_k^-$ there is one point of $\hat \G$ denoted by $\hat P_k^-$
in the neighborhood of which $w$ has zero of order $N$, i.e.
\beq\label{w1}
w=(z-z\left(P_k^-\right))^{N}\left(c_k^-+O(z-z(P_k^-))\right).
\eeq
Let us fix a normalization of the Floque-Bloch solution
by the condition that the sum of coordinates $\psi_0^i$
of the vector $\psi_0$ equals $1$,
\beq\label{nor}
\sum_{i=1}^r \psi_0^i=1.
\eeq
Then, the corresponding Floque-Bloch solution $\psi_n(Q)$ is well-defined
for each point $Q$ of $\hat \G$.
\begin{theo}
The vector-function $\psi_n(Q)$ is a meromorphic vector-function on $\hat\G$,
such that: (i)
outside the punctures $\hat P_{k}^{\pm}$ (which are the points of
$\hat \G$ situated on marked sheets over $P_k^{\pm}$) the divisor $\hat \g$ of
its poles $\hat \g_{\s}$ is {\it $n$-independent}; (ii) at the punctures
$\hat P_k^{+}$ and $\hat P_k^-$ the vector-function $\psi_n(Q)$ has poles and zeros
of the order $n$, respectively; (iii) in the general position, when
$\hat \G$ is smooth, the number of these poles equals $\hat g+r-1$, where
$\hat g$ is the genus of $\hat \G$.
\end{theo}
{\it Proof.} The coordinates of the vector-function $\psi_0(Q)$ are rational
expressions in $w$ and the entries of $T_0$. Hence, it is a meromorphic function
on $\hat \G$. Let $\hat \g_{\s}$ be a set of the poles of $\psi_0$.
In order to show that $\hat \g_{\s}$ are the only poles of
$\psi_n=L_{n-1}\cdots L_0 \psi_0$ outside the preimages on $\hat \G$
of the punctures $P_k^+$, it is enough to prove by induction that
$\psi_n$ at all the preimages $\g_s^i(n)$ of the points $\g_s(n)$ satisfies
the equation $\a_s(n)\psi_n(\g_s^i(n))=0$.  The step of the induction
is a direct corollary of (\ref{Ls1}). The initial statement of the induction
follows from the equation $T_0\psi_0=w\psi_0$. Indeed, $w$ is regular at $\g_s^i(0)$.
Therefore, the left hand side of the equation
has to be regular at these points as well.
The monodromy matrix $T_0$ has a simple pole at $\g_s(0)$ of the form
$m_s\a_s^T(0)$, where $m_s$ is some $r$-dimensional vector. Hence,
$\a_s(0)\psi_0(\g_s^i(0))=0.$
The proof of $(ii)$ is based on the following statement.
\begin{lem}
Let $\tilde L_n$ be a formal series of the form
\beq\label{ll}
\tilde L_n=h(n)p(n)^T\lambda^{-1}+\sum_{i=0}^{\infty}\chi_i(n)\lambda^i
\eeq
where $q(n),p(n)$ are vectors and $\chi_i(n)$ are matrices. Then the equations
\beq\label{ll0}
\phi_{n+1}=\tilde L_n\phi_n,\ \ \phi_{n+1}^*\tilde L_n=\phi_n^*,
\eeq
where $\phi_n$ and $\phi_n^*$ are $r$-dimensional vectors and co-vectors
over the field of the Laurent series in the variable $\lambda$, have
$(r-1)$-dimensional spaces of solutions of the form
\beq\label{ll3}
\Phi_n=\sum_{i=0}\tilde\xi_i(n)\lambda^i,\ \
\Phi_n^*=\sum_{i=0}\tilde\xi_i^*(n)\lambda^i.
\eeq
The equations (\ref{ll}) have  unique formal solutions of the form
\beq\label{ll1}
\phi_n=\lambda^{-n}\left(\sum_{i=0}^{\infty}\xi_i(n)\lambda^i\right),\
\phi_n^*=\lambda^{n}\left(\sum_{i=0}^{\infty}\xi_i^*(n)\lambda^i\right),
\eeq
such that
\beq\label{ll11}
\left(\Phi_n^*\phi_n\right)=\left(\phi_n^*\Phi_n\right)=0,\ \
\left(\phi_n^*\phi_n\right)=1,
\eeq
and normalized by the conditions
\beq\label{ll2}
\chi_0(0)=q(-1), \ \ \sum_{j=1}^i\chi_i^j(0)=0,\ i>0.
\eeq
\end{lem}
For the proof of the lemma it is enough to substitute the formal series (\ref{ll3})
or (\ref{ll1}) in (\ref{ll0}) and use recurrent relations for the coefficients
of the Laurent series.

The subspaces of the solutions of equation (\ref{ll0}) of the form (\ref{ll3})
are invariant under the monodromy operator. Therefore,
euqation (\ref{w}) and the uniqueness of the formal
solutions (\ref{ll1})  imply that
the Laurent expansion of the Floque-Bloch solution $\psi_n$ in the neighborhood
of $\hat P_k^{+}$ has the form (\ref{ll1}), where $\lambda=(z-z(P_k^+))$.
Hence, $\psi_n$ has the pole of order $n$ at $\hat P_k^+$.
The same arguments used for the equation $\psi_n=L_n^{-1}\psi_{n+1}$ imply that
$\psi_n$ has zero of order $n$ at the punctures $\hat P_k^-$.

Let $S^{D_+,D_-}$ be a space of the meromorphic functions $r_i(z)$ on $\G$
with poles of order $N$ at the punctures $P_k^+$, and such that $r_0$ has zeros
of order $N$ at the punctures $P_k^-$.
The Riemann-Roch theorem implies that $S^{D_+,D_-}$ is of dimension
\beq\label{dimS}
\dim \ S^{D_+,D_-}=N\D(r-1)-(g-1)(r-1)+1.
\eeq
The characteristic equation (\ref{char}) defines a map $\LL_N\to S^{D_+,D_-}$.
Usual arguments show that this map on an open set is surjective.
These arguments are based on solution of {\it the inverse spectral problem},
which reconstruct $L_n$, modulo gauge equivalence
\beq\label{gauge}
L_n'=g_{n+1}L_ng_n^{-1}, \ \ g_n\in GL_r,
\eeq
from a generic set of spectral data: a smooth curve $\wh \G$ defined
by $\{r_i\}\in S^{D_+,D_-}$, and a point of the Jacobian $J(\wh \G)$, i.e.
the equivalence class $[\wh \g]$
of degree $\wh g+r-1$ divisor $\wh \g$ on $\wh \G$.
Here $\wh g$ is the genus of $\wh \G$.

For a generic point of $S^{D_+,D_-}$ the corresponding spectral curve $\wh\G$
is smooth. Its genus $\wh g$ can be found with the help of the Riemann-Hurwitz
formula $2\wh g-2=2r(g-1)+\deg\nu$, where $\nu$ is the  divisor on $\G$, which  is
projection of the branch points of $\wh \G$ over $\G$. The branch points
are zeros on $\wh \G$ of the function $\p_wR(w,z)$.
This function has the poles of order $N(r-1)$ on the marked sheet over $P_k^+$,
and poles of order $N$ on all the other sheets. The numbers of poles and
zeros of a meromorphic function are equal. Therefore, $\deg\nu=2N\D(r-1)$ and we
obtain
\beq\label{genus}
\wh g=N\D(r-1)+r(g-1)+1.
\eeq
Moreover, the product of $\p_k R$ on all the sheets of
$\wh \G$ is a meromorphic function on $\G$. Its divisor of
zeros coincides with $\nu$ and the divisor of poles is $N(r-1)D_+$. Therefore,
these divisors are equivalent, i.e.
in the Jacobian $J(\G)$ of $\G$ we have the equality
\beq\label{j}
[\nu]=2N(r-1)[D_+]\in J(\G).
\eeq
The degree of the divisor $\wh \g$ of the poles of $\psi_0$
can be found in the usual way. Let $\Psi_n(q), q\in \G,$ be a matrix with columns
$\psi_n(Q^i)$,
where $Q^i=(w_i(q),q)$ are preimages of $q$ on $\wh \G$
\beq\label{p1}
\Psi_n(q)=\{\psi_n(Q^1),\ldots,\psi_n(Q^r)\}.
\eeq
This matrix depends on ordering of the roots $w_i(q)$ of (\ref{char}),
but the function $F(q)=\det^2\Psi_0(q)$ is independent of this. Therefore, $F$
is a meromorphic function on $\G$.
Its divisor of poles equals $2 \pi_*( \wh \g)$, where $\pi:\wh \G\to \G$
is the projection. In the general position, when the branch
points of $\wh \G$ over $\G$ are simple, the function $F$ has simple zeros at
the images of the branch points, and double zeros at the points $\g_s(0)$, because
evaluations of $\psi_0$ at preimages of $\g_s$ span the subspace orthogonal
to $\a_s(0)$. Therefore, the zero divisor of $F$ is $\nu+2\g(0)$, where $\g(0)=
\g_1(0)+\cdots+\g_{rg}(0)$, and we obtain the equality for equivalence classes
of the divisors
\beq\label{j2}
2[\pi_*(\wh \g)]=[\nu]+2[\g(0)]=2[\g(0)]+2N(r-1)[D_+],
\eeq
which implies
\beq\label{j3}
\deg \wh \g=\deg\nu/2+rg=\wh g+r-1.
\eeq
The theorem is proven.

Let us fix a point $P_0$ on $\G$, and let $\Psi_n$ be the matrix defined by
(\ref{p1}) for $q=P_0$.
Normalization (\ref{nor}) implies that $\Psi_0$ leaves the co-vector
$e_0=(1,\ldots,1)$ invariant, i.e.
\beq\label{nor1}
e_0\Psi_0=e_0.
\eeq
The spectral curve $\wh \G$ and the pole divisor $\wh \g$ are invariant
under the gauge transformation $L_n\to \Psi_{n+1}^{-1}L_n \Psi_n,
\psi_n\to \Psi_n^{-1}\psi$, but the matrix $\Psi_n$ gets transformed
to the identity $\Psi_n=I$. Let $F={\rm diag} (f_1,\ldots, f_r)$ be a
diagonal matrix, then the gauge transformation
\beq
L_n\to FL_nF^{-1},\ \ \psi_n(Q)\to f^{-1}(Q)F\psi_n, \ \ {\rm where} \ \
f(Q)=\sum_{i=1}^r f_i\psi_i(Q),
\eeq
which preserves the normalization (\ref{nor}) and the equality $\Psi_n=I$,
changes $\wh \g$ to an equivalent divisor $\wh \g'$ of zeros of the meromorphic
function $f(Q)$. The gauge transformation of $L_n$ by a permutation matrix
corresponds to the permutation of preimages $P_0^i\in \wh \G$ of $P_0\in \G$,
which was used to define $\Psi_0$.

A matrix $g$ with different eigenvalues has  representation of the form
$g=\Psi_0F$, where $\Psi_0$ satisfy (\ref{nor1}) and $F$ is a diagonal matrix.
That representation is unique up to conjugation by
a permutation matrix. Therefore, the correspondence described above
$L_n\to \{\wh \G,\wh \g, \Psi_n\}$ descends
to a map
\beq\label{smap}
\LL_N/GL_r^N\longmapsto \{\wh \G, [\wh \g]\},
\eeq
which is well-defined on an open set of $\LL_N/GL_r^N$.

According to the Riemann-Roch theorem for each smooth genus $\wh g$
algebraic curve $\wh \G$ with fixed points $q^1,\ldots, q^r, \hat P_k^{\pm}$, and
for each nonspecial degree $\wh g+r-1$ effective divisor $\wh \g$, there
is a unique meromorphic function $\psi_n^i(Q), Q\in \wh \G$, such that:
$\psi_n^i$ has poles of order $n$ at $P_k^+$, and zeros of order $n$ at $P_k^-$;
outside these points it has divisor of poles in $\wh \g$; $\psi_n^i$ is normalized
by the conditions $\psi_n^i(q^j)=\delta_i^j$. Let $\psi_n(Q)$ be a meromorphic vector-function
with the coordinates $\psi_n^i(Q)$. Note, that it satisfies (\ref{nor}).

Let $\wh \G$ be a smooth algebraic curve that is an $r$-fold branch cover of $\G$
$\pi:\hat \G\to \G$. Then for each point $q\in \G$ we define the matrix $\Psi_n(q)$
with the help of (\ref{p1}). It depends on a choice of order of the sheets of the cover $\pi$,
but the matrix function
\beq\label{inv}
\tilde L_n(q)=\Psi_{n+1}(q)\Psi_n^{-1}(q),
\eeq
is independent of the choice, and therefore,
is a meromorphic matrix function on $\G$. It has simple poles at
$P_k^+\in D$ and is holomorphic at the points of the branch divisor $\nu$.
By reversing the arguments used for the proof of (\ref{j3}), we get that the
degree of the zero divisor $\g$ of $\det \Psi_n$ equals $rg$. In general
position the zeros $\g_s(n)$ are simple. The
expansion of $\tilde L_n$ at $\g_s(n)$ satisfies constraints
(\ref{Ls0},\ref{Ls1}), where $\a_s(n)$ is a unique (up to multiplication)
vector orthogonal to the vector-columns of $\Psi_n(\g_s(n))$.
The determinant of $\Psi_n$ has zeros of order $n$ at the points $P_k^-$.
Hence, the determinant of $L_n$ has simple zeros at $P_k^-$.
Therefore, $\tilde L_n\in \LL$.

If $\wh \G$ is defined by equation (\ref{char}), where $(r_j)$
corresponds to a generic point of the space $S^{D_+,D_-}$, and the points $\hat
P_k^{\pm}$ used in the definition of $\psi_n$ are the punctures, at which
$w(Q)$ has poles and zeros of order $N$, then the uniqueness of $\psi_n$ implies
\beq\label{per}
W\psi_{n+N}=w\psi_n,\ \ W^{ij}=w(q^i)\delta^{ij}.
\eeq
From (\ref{per})
it follows that $\tilde L_n=W\tilde L_{n+N}W^{-1}$, and the gauge equivalent
chain $L_n=W^{(n+1)/N}\tilde L_n W^{-n/N}$ is $N$-periodic, $L_n=L_{n+N}$.
If the points $P_0^i$ used for normalization of $\psi_j$ are preimages of
$P_0\in \G$, then $L$, given by (\ref{inv}), is diagonal at $q=P_0$, and
the correspondence $\{\wh \G,\wh \g\}\to L$
descends to a map
\beq\label{smap1}
\{\wh \G,[\wh \g]\}\to \LL_N/GL_r^N,
\eeq
which is well-defined on an open set of the Jacobain bundle over $S^{D_+,D_-}$,
where it is inverse to (\ref{smap}).

\noindent{\bf Restricted chains.} Let us introduce subspaces
$\LLL\subset \LL_N$ of the Lax chains with {\it fixed} equivalence classes
of the divisors of Tyurin parameters
\beq\label{fix}
[\g(n)]=C+n\left([D_+]-[D_-]\right)\in J(\G),
\eeq
and with {\it fixed} determinant $\det T=\Delta=r_0(q)$.
The subspace of the corresponding spectral curves will be denoted by
$\SP_{\Delta}\in \SP^{D_+,\, D_-}$. The points of $\SP_{\Delta}$ are sets of functions $r_i(q),\
i=1,\ldots, r-1,$ with the poles of order $N$ at $P_k^+$.
From equation (\ref{j2}) it follows that for the restricted chains
the equivalence class $[\hat \g]\in J(\hat \G)$ of the poles of the Floque-Bloch
solutions belongs to the abelian
subvariety
\beq\label{subv}
J_C(\hat \G)=\pi_*^{-1}\left( C+N(r-1)[D_+]/2\right),\ \
\pi_*: J(\hat \G)\longmapsto J(\G).
\eeq
\begin{cor} The correspondence
\beq\label{maps2}
\LLL/GL_r^N\leftrightarrow \{\hat \G\in \SP_{\Delta}, [\hat \g]\in J_C(\hat \G)\}
\eeq
is one-to one on the open sets.
\end{cor}

\section{Lax equations}

Our next goal is to construct a hierarchy of commuting flows on an open set
of $\LL$. Let us identify the tangent space $T_L(\LL)$ at the point $L=\{L_n\}$
with the space of meromorphic matrix functions spanned by derivatives
$\p_{\tau}L_n|_{\tau=0}$ of all the one-parametric deformations
$L_n(q,\tau)\in \LL$ of $L_n$. Let us show that the latter space
can be identified with the space of matrix functions
$l_n(q)$ on $\G$ such that:

$1^0.$ $l_n$ has simple poles at the punctures $P_k^+$ of the form
\beq\label{T0}
l_n={\dot h_k(n)p_k^T(n)+h_k(n)\dot p_k^T(n)\over z-z(P_k^+)}+O(1),
\eeq
where $\dot h_k(n), \dot p_k(n)$ are vectors, defined up to the gauge
transformation
\beq\label{gg1}
\dot h_k(n)\longmapsto \dot h_k(n)+\tilde c_{n,k}h_k(n),\ \
\dot p_k(n)\longmapsto \dot p_k(n)-\tilde c_{n,k}p_k(n).
\eeq
The vectors $h_k(n), p_k(n)$ are defined by the expansion (\ref{t1})
of $L_n$.

$2^0.$ $l_n$ has double poles at the points $\g_s$, where it has the expansion of
the form
\beq\label{T}
l_n=\dot z_s(n){\b_s(n)\a_s^T(n)\over (z-z_s(n))^2}+
{\dot\b_s(n)\a_s^T(n)+\b_s(n)\dot\a_s^T(n)\over z-z_s(n)}+
O(1).
\eeq
Here $\dot z_s(n)$ is a constant, and $\dot \a_s(n), \ \dot\b_s(n)$ are
certain vectors.
The vectors $\a_s(n),\b_s(n)$ are defined by $L_n$.

$3^0.$ In addition it is required that the following equation holds:
\beq\label{T1}
\a_s(n+1)l_n(\g_s(n+1))+\dot\a_s(n+1)L_n(\g(n+1))+\dot z_s\a_s(n+1)L'_n(\g_s(n+1))=0,
\eeq
where $L'=\p_zL(z)$.

$4^0.$ The function
\beq\label{T2}
{\rm Tr}\ \left(l_nL_n^{-1}\right)=O(1),\ \ z\to P_k^-
\eeq
is regular at the punctures $P_k^-$.

The constraints ($1^0-4^0$) can be easily checked for a tangent vector
$\p_{\tau}L|_{\tau=0}$, if we identify $(\dot z_s, \dot\a_s, \dot\b_s, \dot p_k,
\dot q_k)$ with the derivatives
\beq
\dot z_s=\p_{\tau} z(\g_s(\tau)),\
\dot \a_s=\p_{\tau} \a_s(\tau)),\
\dot \b_s=\p_{\tau} \b_s(\tau)),\
\dot h_k=\p_{\tau} h_k(\tau)),\
\dot p_k=\p_{\tau} p_k(\tau)).
\eeq
Direct counting of the number of the constraints shows that
the space of matrix functions, which
satisfy (\ref{T0}-\ref{T2}), equals the dimension of
$\LL$. Therefore, on an open set these relations are necessary and sufficient
conditions for $l_n$ to be a tangent vector.

\begin{lem} Let $M_n$ be a meromorphic matrix function on $\G$ with poles at $P_k^+$
and with simple poles at $\g_s(n)$ of the form:
\beq\label{eq1}
M_n={\mu_s(n)\a_s^T(n)\over z-z_s(n)}+m_{s}(n)+O(z-z_s), \ z_s(n)=z(\g_s(n)),
\eeq
where $\mu_s(n)$ is a vector.
Then the matrix-function $M_{n+1}L_n-L_nM_n$ is a tangent vector to $\LL$ at $L_n$,
if and only if it has the form (\ref{T0}) in the neighborhood of $P_k^+ $ .
\end{lem}
{\it Proof.} It is straightforward to check that, if we define $\dot z_s(n),
\dot \a_s(n)$ by the formulae (2.7), (2.8) in \cite{kr-hit}, i.e.
\begin{eqnarray}
\dot z_s(n)&=&-\a_s^T(n)\mu_s(n), \ \ z_s=z(\g_s),\label{ttt1}\\
\dot \a_s^T(n)&=&-\a_s^T(n) m_{s}(n),  \label{ttt2}
\end{eqnarray}
then $M_{n+1}L_n-L_nM_n$ satisfies the constraints (\ref{T}) and (\ref{T1}).
The constraint (\ref{T2}) is also satisfied, because
${\rm Tr}\ \left((M_{n+1}L_n-L_nM_n)L_n^{-1}\right)=
{\rm Tr}\ \left(M_{n+1}-M_n)\right)$, and $M_n$ is regular at $P_k^-$.

The Lemma directly implies, that the Lax equation $\p_tL_n=M_{n+1}L_n-L_nM_n$ is
a well-defined system on an open set of $\LL$, whenever we can define
$M_n=M_n(L)$, as a function of $L$, that satisfies the conditions of Lemma 3.1.
Our next goal is to define a set of such functions $M_n^{(\pm k,l)}(L)$,
parameterized by the puncture $P_k^{\pm}$ and a non-negative integer $l\in Z_+$.

Let us fix a point $P_0\in \G$ and local coordinates $z$ in the neighborhoods
of the punctures $P_k^+$. Let $\phi_n^{(k)}, \phi_n^{(*,k)}$ be the formal solutions
in the neighborhood of $P_k^+$
of equation (\ref{laxL}) and the dual equation $\psi_{n+1}^*L_n=\psi_n^*$, which have
the form (\ref{ll1}), where $\lambda=z-z(P_k^+)$.
From Lemma 2.1, applied to the inverse chain $L_n^{-1}$, it follows that
equation
(\ref{laxL}) and its dual in the neighborhood of $P_k^-$  have formal solutions
$\phi_n^{(-k)}, \phi_n^{(*,-k)}$ of the form
\beq\label{lla}
\phi_n^-=\lambda^{n}\left(\sum_{i=0}^{\infty}\xi^-_i(n)\lambda^i\right),\
\phi_n^{*,-}=\lambda^{-n}\left(\sum_{i=0}^{\infty}\xi_i^{*,-}(n)\lambda^i\right).
\eeq
From the Riemann-Roch theorem (see details in \cite{kr-hit})
it follows that there is a unique matrix function
$M_n^{(\pm k,l)}$ such that:

(i)  it has the form (\ref{eq1})  at the points $\g_s$;

(ii) outside of the divisor $\g$ it has pole at the point $P_k^{\pm}$, only, where
\beq\label{in}
M_n^{(\pm k,l)}=\left(z-z(P_k^{\pm})\right)^{-l}\phi_n^{(\pm k)}\phi_n^{(*,\pm k)}+O(1);
\eeq

(iii) $M_n^{(\pm k,l)}$ is normalized by the condition $M_n^{(\pm k,l)}(P_0)=0$.

\noindent
Note, that although $\phi_n^{(\pm k)}$ and $\phi_n^{(*,\pm k)}$ are formal series,
the constraint (\ref{in}) involves only a finite number of their coefficients, and
therefore, is well-defined.
\begin{theo} The equations
\beq\label{La}
\p_a L_n=M_{n+1}^aL_n-L_nM_n^a ,\ \ \p_a=\p/\p t_a,\ \ a=(\pm k,l),
\eeq
define a hierarchy of  commuting flows on an open set of $\LL$, which
descents to the commuting hierarchy on an open set of $\LL_N/GL_r^N$.
\end{theo}
Equation (\ref{in}) implies that $M_n^a$ satisfies the conditions of
Lemma 3.1. Therefore, the right hand side of (\ref{La}) is a tangent vector
to $\LL$ at the point $L$. Hence, (\ref{La}) is a well-defined dynamical
system on an open set of $\LL$.
Commutativity of flows (\ref{La}) is equivalent to the equation
\beq\label{com}
\p_bM_n^a-\p_aM_n^b+
[M_n^a,M_n^b]=0.
\eeq
As shown in the proof of Theorem 2.1 of \cite{kr-hit}, this equation holds
if its left hand side is regular at the points $P_k^{\pm}$.
From the uniqueness of the formal solutions $\phi_n^{(\pm k)}, \
\phi_n^{(*,\pm k)}$ it follows that
\begin{eqnarray}
\p_a\ \phi_n^{(\pm k)}&=&M_n^{a}\phi_n^{(\pm k)}-f^{(\pm k,\, a)}\phi_n^{(\pm k)},
\label{M}\\
- \p_{a}\ \phi_n^{(*,\pm k)}&=&M_n^{a}\phi_n^{(*,\pm k)}-f^{(\pm k,\,  a)}
\phi_n^{(*,\pm k)},
\label{M1}
\end{eqnarray}
where $f^{(\pm k, a)}$ are scalar functions. The left hand side of (\ref{M})
is regular at all the punctures $P_m^{\pm}$. Vanishing of the singular terms
of the right hand side of these equations implies that for $a=(\pm m,l)$
\beq\label{lf}
f^{(\pm k, \, a)}=\delta_{m,k}(z-z(P_k^{\pm})^{-l}+O(1).
\eeq
Equations (\ref{M},\ref{M1}) and standard arguments used in  KP theory
(see details in \cite{kr-hit})) imply, that
the left hand side of (\ref{com}) is regular at $P_k^{\pm}$.
By definition, the matrix functions $M_n^a$ are periodic, if $L_n$ are
periodic.
The matrices $M_n^a$ under the transformations (\ref{gauge}) get transformed
to $\tilde M_n^a=g_nM_n^ag_n^{-1}$.
Therefore, the flows (\ref{La}) are well-defined on $\LL_N$.
The theorem thus is proven.

In general the flows (\ref{La}) do not preserve the leaves of the
foliation $\LLL\subset \LL_N$. Let us find their linear combinations for which
the subspaces of the restricted chains are invariant.
Let $f$ be a meromorphic function on $\G$ with poles only
at the punctures $P_k^{\pm}$. Then we define
\beq\label{tf}
M_n^f=\sum_a c_{a}^f M_n^a,
\eeq
where $c_{a}^f$ are the coefficients of the singular part of the Laurent expansion
\beq\label{tf1}
f=\sum_{l>0} c_{(\pm k,l)}^f \left(z-z(P_k^{\pm})\right)^{-l} + O(1).
\eeq
\begin{theo} The equations
\beq\label{Lf}
\p_f L_n=M_{n+1}^fL_n-L_nM_n^f ,\ \ \p_f=\p/\p t_f,
\eeq
define a hierarchy of  commuting flows on an open set of $\LLL$, which
descents to the commuting hierarchy on an open set of
$\LLL/GL_r^N$.
\end{theo}
{\it Proof.} The flows (\ref{Lf}) are linear combinations with constant coefficients
of the basic flows (\ref{La}). Therefore, they are well-defined and commute with
each other. Let us consider the common Floque-Bloch solution $\hat \psi_n(t,Q)$
of (\ref{laxL}) and the equation
\beq\label{laxM}
\p_f\hat \psi_n=M_n^f \hat \psi_n,
\eeq
normalized by the condition (\ref{nor}) at $t=0$. Then equations (\ref{M}-\ref{tf1})
imply that in the neighborhood of $P_k^{\pm}$ the function $\hat\psi_0(t,Q)$
has the form
\beq\label{ff}
\hat\psi_0=e^{t_ff(z)} O(1).
\eeq
Standard arguments, used in the construction the Baker-Akhiezer functions,
imply that outside the punctures $\hat P_k^{\pm}$ the functions $\hat\psi_n$
has {\it time-independent} poles at the pole divisor
$\hat \g(0)$ of $\hat \psi_0$. The pole divisor
$\hat \g(t_f)$ of $\psi_n(t_f,Q)$ is the divisor of zeros of the function
$F(t,Q)=\sum_i \psi_0^i(t,Q),$
where $\psi_0^i$ are the coordinates of $\psi_0$. The function
$\tilde F(t,q)=\prod_j F\left(t,Q^i(q)\right),$
where $Q^i$ are the preimages of $q$ on $\wh \G$, is
meromorphic on $\G$ outside $P_k^{\pm}$. Equation (\ref{ff}) implies that
$\tilde F(t_f,q)e^{-t_f f(q)}$ is a meromorphic function on $\G$.
It has poles at the divisor $\pi_* (\hat \g(0))$ and zeros at the divisor
$\pi_*(\hat \g(t_f))$. Therefore, these divisors are equivalent and
the Theorem is thus proven.

\section{Hamiltonian approach}
In this section we apply the general algebraic approach to the Hamiltonian
theory of the Lax equations proposed in \cite{kp1,kp2}, and developed in
\cite{kr4}, to the Lax equations for {\it periodic} chains on the algebraic curves.
As it was mentioned in the introduction, this approach is based on the existence of
two universal two-forms on a space of meromorphic matrix-function. They
can be traced back to the fact that there are two basic algebraic structures on
a space of {\it operators} (see details in \cite{kp3}) .
The first one is the Lie algebra structure defined by the commutator of operators.
The second one is the Lie group structure. The discrete Lax equations or chains
correspond to the Lie group structure.

The entries of $L_n(q)\in \LL_N$ can be regarded as functions on $\LL_N$. Therefore,
$L_n$ by itself can be seen as matrix-valued function and its external derivative
$\delta L_n$ as a matrix-valued one-form on $\LL_N$. The matrix $\Psi_n$ (\ref{p1})
with columns formed by the canonically normalized Floque-Bloch solutions
$\psi_n(Q^i)$ of (\ref{laxL}) can also be regarded as a matrix function on $\LL_N$
(modulo permutation of the columns). Hence, its differential $\delta \Psi_n$
is a matrix-valued one-form on $\LL_N$.
Let us define a two-form $\Omega(z)$ on $\LL_N$ with values in the space
of meromorphic functions on $\G$ by the formula
\beq\label{Om}
\Omega(z)=
\sum_{n=0}^{N-1}{\rm Tr} \left(\Psi_{n+1}^{-1} \delta L_n\wedge \delta \Psi_n\right).
\eeq
It can be also represented in the form
\beq\label{Om11}
\Omega(z)={\rm Tr} \left(\Psi_{N}^{-1} \delta T\wedge \delta \Psi_0\right)=
{\rm Tr} \left(\Psi_{0}^{-1} T^{-1}\delta T\wedge \delta \Psi_0\right)=
{\rm Tr} \left(W^{-1}\Psi_{0}^{-1} \delta T\wedge \delta \Psi_0\right),
\eeq
where $W={\rm diag }(w_i(z))$ is a diagonal matrix, whose diagonal entries
are the eigenvalues of the monodromy matrix $T=T_0$

Fix a meromorphic differential $dz$ on $\G$ with poles at a set of points
$q_m$. Then the formula
\beq\label{form}
\omega=-{1\over 2}\sum_{q\in I}\res_q \Omega \, dz, \ \  I=\{\g_s,P_k^{\pm}, q_m\}
\eeq
defines a scalar-valued two-form on $\LL_N$. This form depends on a choice of
the normalization of $\Psi_n$. A change of the normalization
corresponds to the transformation $\Psi_n'=\Psi_n V$, where $V=V(z)$ is
a diagonal matrix, which might depend on a point $z$ of $\G$.
The corresponding transformation of $\Omega$ has the form:
\beq\label{trans1}
\Omega'=\Omega+\delta \left({\rm Tr}\left(\ln W v\right)\right),\ \
v=\delta V V^{-1}.
\eeq
Here we use the standard formula for a variation
of the eigenvalues
\beq\label{Om1}
\Psi_0^{-1}\delta T\Psi_0=\delta W +\Psi_0^{-1}\delta \Psi_0 W -
W \Psi_0^{-1}\delta \Psi_0,
\eeq
and the equation $\delta v=v\wedge v=0$, which is valid
because $v$ is diagonal.

Let $\X^{D_+,\, D_-}$ be a subspace of the chains $\LL_N$ such,
that the restriction of $\delta (\ln w)\, dz$ to $\X^{D_+,D_-}$  is a differential
{\it holomorphic} at all the preimages on $\hat \G$ of the punctures $P_k^{\pm}$.
\begin{lem} The two-form $\omega$, defined by (\ref{form}) and restricted to
$\X^{D_+,\,D_-}\subset \LL_N$, is independent of the choice
of normalization of the Floque-Bloch solutions, and is gauge invariant, i.e.
it descends to a form on
$\P=\X^{D_+,\, D_-}/GL_r^N$.
\end{lem}
The proof of the lemma is almost identical to the proof of Lemma 2.4
in \cite{kr-hit}.

By definition, a vector field $\p_t$ on a symplectic
manifold is Hamiltonian, if the contraction $i_{\p_t}\omega(X)=
\omega(\p_t,X)$ of the symplectic form
is an exact one-form $dH(X)$. The function $H$ is the Hamiltonian,
corresponding to the vector field $\p_t$. The proof of the following theorem
is almost identical to the proof of Theorem 4.2 in \cite{kr-hit}.
\begin{theo}
Let $\p_a$ be the vector fields corresponding to the Lax equations (\ref{La}).
Then the contraction of $\omega$, defined by (\ref{form}) and
restricted to $\P$, equals
\beq\label{H}
i_{\p_a}\omega=\delta H_a,
\eeq
where
\beq\label{H1}
H_{(\pm k,\,l)}=\res_{P_k^{\pm}}\left(z-z\left(P_k^{\pm}\right)\right)^{-l}
\left(\ln w\right)  dz.
\eeq
\end{theo}
The theorem implies that Lax equations (\ref{La}) are Hamiltonian whenever
the form $\omega$ is non-degenerate. In order to analyze this problem we first
find the Darboux variables for $\omega$.
\begin{theo}
Let $L_n\in \LL_N$ be a periodic chain, and let $\wh \g_s$ be the poles of the
normalized (\ref{nor}) Floque-Bloch solution $\psi_n$. Then
the two-form $\omega$ defined by (\ref{form}) is equal to
\beq\label{d1}
\omega=\sum_{s=1}^{\wh g+r-1} \delta \ln w(\wh \g_s)\wedge \delta z(\wh \g_s).
\eeq
\end{theo}
The proof of the Theorem is analogous to the proof of the Theorem 4.3 in
\cite{kr-hit} and equation (5.7) in \cite{kp4}.
The meaning of the right hand side of the formula (\ref{d1}) is as follows.
By definition, the spectral curve is equipped
with the meromorphic function $w(Q)$. The pull-back of
the Abelian integral $z(q)=\int^{q}dz$ on $\G$ is a multi-valued holomorphic
function on $\wh \G$.
The evaluations $w(\wh \g_s),\ z(\wh \g_s)$ at the points
$\wh \g_s$ define functions on the space $\LL_N$, and the wedge product of
their external differentials is a two-form on $\LL_N$. (Note, that
differential $\delta z(\wh \g_s)$ of the multi-valued function
$z(\wh g_s)$ is single-valued, because the periods of $dz$ are constants).

From (\ref{d1}) it follows that $\omega$ can be represented in the form:
\beq\label{form1}
\omega=\sum_{k=1}^{\hat g} \delta A_k\wedge \delta \varphi_k,
\eeq
where $\varphi_k$ are the coordinates on $J(\hat \G)$, corresponding to a choice
of $a$- and $b$-cycles on $\hat \G$ with the canonical matrix of intersections,
and
\beq\label{A}
A_k=\oint_{a_k} (\ln w)\,  dz.
\eeq
Note, that the external differential $\delta A_k$ of the multi-valued function
$A_k$ is single-valued, because all the periods of $dz$ are fixed.

The spectral map (\ref{smap}) identifies an open set of
$\LLL/GL_r^N$ with an open set of the Jacobian bundle over
$\SP_{\Delta}\subset \SP^{D_+,\, D_-}$, i.e.
\beq\label{bundle}
\LLL/GL_r^N\longmapsto \SP_{\Delta}.
\eeq
From (\ref{form1}) it follows, that the form $\omega$ can be non-degenerate
only if the base and the fibers of the bundle (\ref{bundle}), restricted
to $\P$, have the same dimension.

First, let us consider the case of the chains on the rational curve (see details
in \cite{kp3}). The basic example is the chain corresponding to the Toda lattice,
in which $L_n$ has the form
\beq\label{Toda}
L_n=\left(\begin{array}{cc} 0&1\\c_n &z+v_n\end{array}\right).
\eeq
For $g=0$ equations (\ref{dimS}) and (\ref{genus}) imply that
the space of the spectral curves and their Jacobians are of dimensions
$N\D(r-1)+r-1$ and $N\D(r-1)-r+1$, respectively. The differential $dz$ has
double pole at the infinity. Therefore, the subspace
$\X^{D_+,\, D_-}\subset \LL_N$ is defined by the constraint
that the eigenvalues of the monodromy matrix are fixed
up to the order $O(z^{-2})$. The number of corresponding equations on $\SP_{\Delta}$
is $2r-1$.
Let $\P_*$ be a subspace of $\P$ correspodning to the restricted chains. For $g=0$
that means that the determinant of the monodromy matrix is fixed. Then
$\dim \,\P_*=2\, \dim\, J(\hat\G)$ and arguments identical to that used
at the end of Section 4 in \cite{kr-hit} prove, that the form $\omega$ is
non-degenerate on $\P_*$.

Consider now the case $g>0$. Let $dz$ be a holomorphic differential on $\G$. Then,
for each branch of $w=w_i(z)$ the differential $\delta (\ln w)\, dz$
is always holomorphic at $P_k^{\pm}$. Hence, $\P=\LL_N/GL_r^N$.
Recall, that
\beq\label{dimss}
\dim\,  \SP_{\Delta}=(r-1)(N\D-g+1),\ \  \dim\,  J(\hat \G)=(r-1)(N\D+g-1)+g.
\eeq
Therefore, for $g>0$ the form $\omega$ is degenerate on $\P$. For $g=1$ the space
$\P$ is a Poisson manifold with the symplectic leaves, which are factor-spaces
\beq\label{ph}
\P_*=\LLL/GL_r^N
\eeq
of the restricted chains. In that case $\dim\, S_{\Delta}=\dim\, J_C(\hat \G)=N\D(r-1)$.
As in the genus zero case, the arguments identical to that used at the end of Section 4 in
\cite{kr-hit} prove that the form $\omega$ is non-degenerate on $\P_*$.
\begin{cor} For $g=0$ and $g=1$ the form $\omega$ defined by (\ref{form})
descents to the symplectic form on $\P_*$, which coincides with the pull-back
of the Sklyanin symplectic structure restricted
to the space of the monodromy
operators. The Lax equations $(\ref{Lf})$
are Hamiltonian with the Hamiltonians
\beq\label{H2}
H_f=\sum_{q=\hat P_k^{\pm}} \res_{q} (\ln w) f\, dz.
\eeq
The Hamiltonians $H_f$ are in involution $\{H_f, H_h\}=0$.
\end{cor}
The first statement is a direct corollary of equation (\ref{d1}) and
the Proposition 3.33 in \cite{hart}. The second statement is a corollary
of equation (\ref{H1}) and the definition of $H_f$. Finally, involutivity of
the Hamiltonians follows from the commutativity of the corresponding Lax flows.

Now we are in the position to discuss the case $g>1$ mentioned in the Introduction.
The space $\P_*$ is equipped by
$g$-parametric {\it family} of two-forms $\omega_{dz}$, parameterized by
the holomorphic differentials $dz$ on $\G$. For all of them the fibers
$J_C(\hat \G)$ of the spectral bundle are maximum isotropic subspaces.
For each vector-field $\p_f$ defined by (\ref{Lf}) the equation
$i_{\p_f}\omega_{dz}=\delta H_f^{dz}$ holds.

Equation (\ref{form1}) implies, that each of the forms $\omega_{dz}$ is degenerate
on $\P_*$. Let us describe the kernel of $\omega_{dz}$.
According to Theorem 3.2, the tangent vectors to $J_C(\hat \G)$ are parameterized by
the space $\A(\G,P_{\pm})$ of meromorphic functions $f$ on $\G$ with the poles at
$P_k^{\pm}$ modulo the following equivalence relation.
The function $f$ is equivalent to $f_1$, if there is a meromorphic function
$F\in \A(\hat \G,\hat P_k^{\pm})$ on $\hat \G$ with the poles at $\hat P_k^{\pm}$,
such that in the neighborhoods of these punctures the function $\pi^*(f-f_1)-F$
is regular. Let $K_{dz}\subset \A(\G,P_{\pm})$ be the subspace of functions such
that there is a meromorphic function $\tilde F$ on $\hat \G$ with poles
at $\hat P_k^{\pm}$ and at the preimages $\pi^*(q_s),\, dz(q_s)=0$
of the zero-divisor of $dz$, and such that $f-\tilde F$ is regular at
$\hat P_k^{\pm}$. Then, from equations (\ref{H}) and (\ref{H2}) it follows that:
$
f\in K_{dz}\longmapsto i_{\p_f}\omega_{dz}=0.
$
Let $\K_{dz}$ be the factor-space of $K_{dz}$ modulo the equivalence relation.
Then the Riemann gap theorem implies that in the general position $\K_{dz}$
is of dimension $2(g-1)(r-1)$, which equals the dimension of the kernel of
$\omega_{dz}$. Therefore, the kernel of $\omega_{dz}$ is isomorphic to $\K_{dz}$.
Using this isomorphism, it is easy to show that the intersection of all the kernels of the forms
$\omega_{dz}$ is empty, and thus the family of these forms is {\it non-degenerate}.

\section{Variable base curves}

Until now it has been always assumed that the base curve is {\it fixed}. Let
$\M_{\Delta}$ be the space of smooth genus $g$ algebraic curves $\G$ with
the fixed meromorphic function $\Delta$, having poles and zeros of order $N$ at
punctures $P_k^{\pm}, k=1,\ldots, \D$. For simplicity, we will assume that the
punctures $P_k^{\pm}$ are distinct. The space $\M_{\Delta}$ is of dimension
$\dim \, \M_{\Delta}=2(\D+g-1)$. The total space $\LT$ of all the
restricted chains corresponding to these data and the trivial equivalence
class $C=0\in J(\G)$ can be regarded as the bundle
over $\M_{\Delta}$ with the fibers $\LO=\LO(\G)$. By definition, the curve $\G$
corresponding to a point $(\G,\Delta)\in \M_{\Delta}$ is equipped by the meromorphic
differential $dz=d\ln \Delta$. The function $\Delta$ defines
local coordinate everywhere on $\G$ except at zeros of its differential.
Let $\omega_{\Delta}$ be the form defined by (\ref{form}),
where $dz=d\ln \Delta$ and the variations of $L_n$ and $\Psi_n$ are taken
with fixed $\Delta$, i.e.
\beq\label{formd}
\omega_{\Delta}=-{1\over 2}\sum_{q\in I}\res_q
\sum_{n=0}^{N-1}{\rm Tr} \left(\Psi_{n+1}^{-1}(\Delta) \delta L_n(\Delta)\wedge
\delta \Psi_n(\Delta)\right)
 \, d\ln \Delta, \ \  I=\{\g_s,P_k^{\pm}\}.
\eeq
Then $\omega_{\Delta}$ is well-defined on leaves $\hat \X_{\Delta}$
of the foliation on $\LT$ defined by the condition: the differential
$\delta \ln w(\Delta)d\ln \Delta$ restricted to $\hat \X_{\Delta}$ is {\it holomorphic}
at the punctures $P_k^{\pm}$. This condition is equivalent to the following
constraints. In the neighborhood of $P_k^{\pm}$
there are $(r-1)$ regular branches $w_i^{\pm}$ of the multi-valued function $w$,
defined by the characteristic equation (\ref{char}):
\beq\label{v1}
w_i^{\pm}=c_i^{\pm}+O(\Delta^{\mp 1}), \ i=1,\ldots, r-1.
\eeq
The leaves $\hat \X_{\Delta}$ are defined by $2\D (r-1)$ constraints:
\beq\label{v2}
\delta c_i^{\pm}=0\longmapsto c_i^{\pm}=const_i^{\pm}.
\eeq
Note, that the differential $\delta \ln w(\Delta)d\ln \Delta$ is regular at
$P_k^{\pm}$ for the singular branches  of $w$, because the coefficients $c^{\pm}$
of the expansions (\ref{w}) and (\ref{w1}) are also fixed due to the equation
$c^{\pm}\prod_{i=1}^{r-1}c_i^{\pm}=1$.

The factor-space $\hat \P_{\Delta}=\hat \X_{\Delta}/GL_r^N $ is of dimension
$\dim\, \hat \P_{\Delta}=2N\D(r-1)-2\D (r-1)+2(\D+g-1)$.
The space $\hat \SP_{\Delta}^{\,0}\subset \hat S_{\Delta}$ of the corresponding
spectral curves is of dimension $\dim\, \hat \SP_{\Delta}^0=
(r-1)(N\D -g+1)-2\D(r-1)+2(\D+g-1)$. The second and the third terms in the last
formulae are equal to the number of the constraints (\ref{v2}) and the dimension
of $\M_{\Delta}$, respectively.

For the case $r=2$ the last formulae imply the match of the dimensions
$\dim\, \hat \P_{\Delta}=2\,\dim\,  \hat \SP_{\Delta}^0$.
For $r=2$ the spectral curves are two-sheet cover of the base curves, and
the fiber of the spectral bundle is the Prim variety
$J_0(\hat \G)=J_{Prim}(\hat \G)$.

\begin{theo} For $r=2$ the form $\omega_{\Delta}$ defined by (\ref{formd}),
and restricted to $\hat \P_{\Delta}$ is non-degenerate.
If $\wh \g_s$ are the poles of the normalized Floque-Bloch solution
$\psi_n$, then
\beq\label{dv1}
\omega_{\Delta}=\sum_{s=1}^{\wh g+r-1} \delta \ln w(\wh \g_s)\wedge
\delta \ln \Delta(\wh \g_s)=\sum_{s=1}^{\wh g+r-1} \delta \ln w(\wh \g_s)\wedge
\delta \ln w(\wh \g_s^{\sigma}),
\eeq
where $\sigma:\hat \G\to \hat \G$ is the involution, which permutes the sheets
of $\hat \G$ over $\G$.

For every function $f\in \A(\G, P_k^{\pm})$ the Lax equations (\ref{Lf}) are Hamiltonian with the Hamiltonains
$$ H_f=\sum_{q=\hat \P_k^{\pm}} \res_q \, (f\ln w) d\ln \Delta.$$
The Hamiltonians $H_f$ are in involution.
Their common level sets are fibers $J_{Prim}(\hat \G)$ of the spectral map.
\end{theo}


\begin{thebibliography}{**}


\bibitem{kn1}
I.M. Krichever, S.P.Novikov, {\it Holomorphic bundles over algebraic curves
and non-linear equations}, Uspekhi Mat. Nauk {\bf 35} (1980) n 6, 47-68.

\bibitem{kr-hit} I.Krichever, {\it Vector bundles and Lax equations on
algebraic curves}, Comm.Math.Physics {\bf 229} (2002), 229-269, hep-th/0108110.

\bibitem{hit} N.Hitchin, {\it Stable bundles and integrable systems},
Duke Math. Journ. {\bf 54} (1) (1987) 91-114.

\bibitem{kn2} I.M. Krichever, S.P.Novikov, {\it
Two-dimensional Toda lattice, commuting difference operators
and holomorphic vector bundles}, Uspekhi Mat. Nauk {\bf 58} (2003) n 3, 51-88,
math-ph/0308019.

\bibitem{zakh}
V.Zakharov, A.Shabat, {\it Integration of non-linear
equations of mathematical physics by the method of the inverse
scattering problem. II}, Funct.Anal. and Appl. 13 (1979), 13-22.

\bibitem{ft}L.D.Faddeev, L.A.Takhtadjan, {\it Hamiltonian methods in the
theory of solitons}, Springer-Verlag, Berlin, 1987.
\bibitem{rs} A.G.Reiman, M.A. Semenov-Tian-Shansky, {\it
Integrable systems II}, chap.2, in
"Dynamical Systems VII, Encyclopaedia of Mathematical Sciences, vol 16,
V.I.Arnold and S.P. Novikov, eds, Springer-Verlag, Berlin 1994.

\bibitem{kp1} I. Krichever and D.H. Phong,
{\it On the integrable geometry of $N=2$ supersymmetric gauge
theories and soliton equations}, J. Differential Geometry
{\bf 45} (1997) 445-485, hep-th/9604199.

\bibitem{kp2} I. Krichever and D.H. Phong,
{\it Symplectic forms in the theory of solitons},
Surveys in Differential Geometry {\bf IV} (1998),
edited by C.L. Terng and K. Uhlenbeck,
239-313, International Press,
hep-th/9708170;

\bibitem{kp3}
E.D'Hoker, I. Krichever and D.H. Phong, {\it
Seiberg-Witten Theory, Symplectic Forms and Hamiltonian Theory of Solitons},
hep-th/0212313.
\bibitem{tyur}
A.Tyurin, {\it Classification of vector bundles over an algebraic curve of
arbitrary genus}, Amer.Math.Soc.Translat., II, Ser 63 (1967), 245-279.
\bibitem{kr2}
I.M. Krichever,  {\it The commutative rings of ordinary differential operators},
Funk.anal. i pril. 12 (1978), n 3, 20-31.

\bibitem{kr4}
I.M. Krichever,
{\it Elliptic solutions to difference non-linear equations and
nested Bethe ansatz}, in: "Calogero-Moser-Sutherland models",
Springer-Verlag, New-York, 1999,
solv-int/9804016.
\bibitem{hart}
J.C. Hurtubise, E. Markman, {\it Surfaces and the Sklyanin bracket},
preprint arXiv: math.AG/0107010
\bibitem{kp4}
I. Krichever and D.H. Phong, {\it Spin chains with twisted monodromy},
Journal of the Inst. Math. Jussieu 1(3) (2002), 477-492,  hep-th/0110098
\end{thebibliography}
\end{document}